\documentclass[a4paper,12pt]{JHEP3} 
\usepackage{graphicx}
\usepackage{dcolumn}
\usepackage{bm}

\newcommand{\hub}{{\cal H}}
\newcommand{\z}{{\cal Z}}

\newcommand{\s}{\sigma}
\newcommand{\p}{{\prime}}
\newcommand{\pp}{{\prime\prime}}
\newcommand{\ppp}{{\prime\prime\prime}}

\title{Generation of Curvature Perturbations with Extra Anisotropic Stress}

\author{
Kazuhiko Kojima\footnote{Present address: (On leave from University of Tokyo) IBM Global Services Japan Solution and Services Company, Sapporo Center Building, 2-2-21F Kita 5-jyo Nishi 6-chome, Chuo-ku, Sapporo, 060-0005, Japan}\\
Department of Astronomy, Graduate school of Science, University of Tokyo,
7-3-1 Hongo, Bunkyo-ku, Tokyo 113-0033, Japan\\
Division of Theoretical Astrophysics, National Astronomical
Observatory, 2-21-1 Osawa, Mitaka, Tokyo 181-8588, Japan\\
Email: \email{kazuhiko@07.alumni.u-tokyo.ac.jp}
}

\author{
Toshitaka Kajino\\
Division of Theoretical Astronomy, National Astronomical Observatory,
2-21-1 Osawa, Mitaka, Tokyo 181-8588, Japan\\
Department of Astronomy, Graduate school of Science, University of Tokyo,
7-3-1 Hongo, Bunkyo-ku, Tokyo 113-0033, Japan\\
Email: \email{kajino@nao.ac.jp}
}

\author{
Grant J. Mathews\\
Center for Astrophysics, Department of Physics, University of Notre Dame, 
Notre Dame, IN 46556, U.S.A.\\
Division of Theoretical Astronomy, National Astronomical Observatory,
2-21-1 Osawa, Mitaka, Tokyo 181-8588, Japan\\
Email: \email{gmathews@nd.edu}
}

\date{\today}
\abstract{
We study the evolution of   curvature perturbations and the cosmic microwave background (CMB) power spectrum in the presence of
an hypothesized extra anisotropic stress which
might arise, for example, from 
the dark radiation term in brane-world cosmology.
We evolve the scalar modes of  such perturbations before and after neutrino decoupling and analyze their effects on the CMB spectrum.  A novel result of this work is that  the cancellation of the neutrino and extra anisotropic stress could lead to a spectrum of residual  curvature perturbations which  is similar to  the observed CMB power spectrum.
This implies a possible additional consideration in the determination of cosmological parameters from the CMB analysis. }
\received{\today} 		
\accepted{\today}	


\begin{document}

\section{Introduction}


Anisotropic stress is the traceless component in the energy-momentum tensor
and it accounts for  anisotropic momentum flow in the universe.
Although  some previous works  have considered  neutrino anisotropic stress and
the anisotropic stress of the primordial magnetic field \cite{1995ApJ...455....7M,2008PhRvD..77f3003G,2006CQGra..23.4991G,2006PhRvD..74f3002G,2008arXiv0806.2018K,2008nuco.confE.226K},
they mainly considered the post neutrino decoupling era.
In this work we deduce 
the evolution of the neutrino anisotropic stress and curvature perturbations both before
 and after neutrino decoupling which can be important.

In the standard cosmological model,
the anisotropic stress is absent before the epoch of neutrino decoupling 
because rapid interactions among elementary particles dissipate it. After the neutrinos have  decoupled
from the cosmic expansion, however, the neutrinos 
become relativistic free particles and neutrino anisotropic stress can grow gradually.
The neutrino anisotropic stress plays an important role in the 
 formation of large scale structure and the CMB fluctuation spectrum.

In this paper, however,  we consider other possible sources for anisotropic stress besides neutrinos or a magnetic field which can be present before neutrino decoupling. 
In particular,  such terms arise naturally in cosmological theories in higher dimensions.  We follow
the evolution of the neutrino anisotropic stress and curvature perturbations both before
 and after neutrino decoupling in the presence of such anisotropic stress terms and show that curvature perturbations can stay constant on superhorizon scales as does the standard adiabatic mode.  
However, unlike the standard case, significant evolution of curvature perturbations occurs before neutrino decoupling.  
We show that, for the right conditions, such curvature perturbations might even reproduce the observed CMB power spectrum implying a possible additional consideration in the extraction of cosmological parameters from the CMB analysis.

One possible source of extra anisotropic stress  is the so-called "dark radiation" term in brane-world cosmology.  This term can 
 affect strongly the CMB anisotropies \cite{2001PhRvD..63h4009L,2002PhRvD..65j4012L,2002PhLB..532..153B,2002PhRvD..66d3521I, 2003PhRvD..68h3518I, 2006PhRvD..73f3527U}.
In such brane-world cosmology, there are  two correction terms in the 4D Einstein
 equation. One is the extrinsic curvature which
 introduces a quadratic term in the  energy momentum tensor, but is negligible in the  low energy limit. The other  is from the projected Weyl tensor. 
The application of the five-dimensional conservation condition to the 
Weyl tensor requires that the energy density term 
varies with scale factor as $a^{-4}$  to an observer on the brane \cite{2001PhRvD..63h4009L}.  Hence, it is
 called the  dark radiation term and remains significant throughout the radiation-dominated epoch.

There is, however,  no intrinsic brane equation to determine the anisotropic stress \cite{2001PhRvD..63h4009L}.   
Although there  have been attempts (e.g. \cite{2003PhRvL..91v1301K}) to constuct simple models for the Weyl anisotropic stress, 
the solution strongly depends upon unknown physics in the bulk dimension.  
Hence, in what follows we take a  phenomenological approach and simply adopt the plausible assumption that in the same way that inflation 
generates scale-free fluctuations characterized by a spectral 
index $P(k) \sim k^n$ (with $n \sim 1$), 
we can expect that the Weyl anisotropic stress could be 
characterized by a spectral index  to be determined (constrained)  
by a fit to the observed CMB power spectrum.
Moreover, this dark-radiation term is assumed to scale  
as $\propto a^{-4}$ similar to the energy density and pressure as was adopted in Ref. \cite{2002PhLB..532..153B}.

Another more familiar  example of extra  anisotropic stress is that of  a primordial magnetic field (PMF).
The amplitude of the 
 energy density $B^2/8\pi$ and magnetic anisotropic stress $\rho_\gamma\pi_B$ 
of the PMF
again both scale as radiation density $\propto a^{-4}$.  Moreover, such a PMF should be characterized by an amplitude and spectral index,
However, the contribution of a PMF to the CMB power spectrum is constrained 
to be small   because it has a measurable effect on the observed BB mode of the CMB on small 
angular scales \cite{2008PhRvD..77d3005Y, 2008PhRvD..78l3001Y}.  
As shown in Ref.~\cite{2004PhRvD..70d3011L}, the tensor mode of 
the primordial magnetic field has two kinds of 
fluctuations. One is a compensated mode, which arises from the compensation of anisotropic stress,
and the other is a passive mode which was generated by the anisotropic stress of the magnetic field before the epoch of neutrino decoupling.
The passive tensor mode makes a large contribution to the CMB.
Although Ref.~\cite{2004PhRvD..70d3011L} also studied the vector mode, there was no analysis of the scalar mode
in that work.
The spectral index of a PMF is also constrained from the effects of the associated gravity waves on  primordial nucleosynthesis \cite{2000PhRvD..61d3001D}.  
Hence, a PMF cannot source the scale-free extra anisotropic stress of interest to this work.  
We only mention it as an example of a power-law anisotropic stress which also scales as $\propto a^{-4}$. 

Here we show that a scale-free  anisotropic stress, if present, could 
imply an additional consideration in the determination of cosmological parameters from the CMB.  
As an extreme example, in this paper we show that an extra anisotropic 
stress could even  lead to a CMB spectrum that agrees with observation.  
Such a source is a bit contrived as it must be scale invariant as naturally
occurs in the inflationary scenario. Thus some kind of mechanism would have to occur 
to produce the desired scale-invariant anisotropic stress.  
Our main point is that the possibility exists that such an 
anisotropic stress  term might be present in the early universe 
and that it would lead to curvature perturbations which could reproduce  
the  observed CMB power spectrum. 
This possibility has not previously been pointed out to our knowledge. 

 A second point which we make here is that this extra source of anisotropic stress may produce non-Gaussian fluctuations depending  upon the nature of the source of anisotropic stress.
The WMAP-5yr analysis, has indicated that there is at least a possibility for  non-Gaussianity although the uncertainty is  
large \cite{2008arXiv0803.0547K}.
The Planck mission should constrain non-Gaussianity with high precision.
If non-Gaussianity were actually observed, it could
 suggest (e.g.~\cite{2004PhR...402..103B}) the need for
a new cosmological paradigm which allows for  non-Gaussianity such as that described here.
At the very least, observational limits on non-Gaussianity could place limits on the hypothesis proposed here, depending upon the source of the extra  anisotropic stress.

\section{Evolution Equations}

To eliminate gauge freedom we study the evolution of the anisotropic stress in covariant 
coordinates within the frame of the cold dark matter (CDM frame).
In this frame, the variables can be defined on a supersurface orthogonal to the
CDM 4-velocity $u_\mu$. 
Then, using the notation of Ref.\cite{2002PhRvD..65j4012L}, the anisotropic expansion rate (shear) $\sigma$ and
the inhomogeneity of the expansion $\z$ can be determined from the covariant derivative of
$u_\mu$. We can neglect vorticity in the scalar mode.
We also introduce the Weyl tensor, i.e. traceless part 
of the Riemann curvature tensor, which 
vanishes in the FRW background spacetime.
The magnetic part of the Weyl tensor is negligible in the scalar mode.
Therefore,  we can define the electric part of the Weyl tensor as $\Phi$.
By linearizing the Bianchi and Ricci identities, one can obtain the following five
equations  
for $\Phi$, $\s$ and $\z$ \cite{2002PhRvD..65j4012L,1992ApJ...395...54D,1999ApJ...513....1C}:
\begin{eqnarray}
&&\Phi^\p+\hub\Phi+\frac{1}{2k}\kappa\rho a^2(\tilde{\gamma}\sigma+R_\gamma q_\gamma+R_\nu q_\nu+ R_b q_b+R_\gamma q_{\rm ex})\nonumber\\
&&+\frac{\hub}{2k^2}\kappa\rho a^2(3\tilde{\gamma}-1)(R_\gamma\pi_\gamma+R_\nu\pi_\nu+R_\gamma\pi_{\rm ex})\nonumber \\
&&-\frac{1}{2k^2}\kappa\rho a^2(R_\gamma\pi_\gamma^\p+R_\nu\pi_\nu^\p+R_\gamma\pi_{\rm ex}^\p)=0~,\label{1}\\
&&\sigma^\p+\hub\sigma+k\Phi+\frac{1}{2k}\kappa \rho a^2(R_\gamma \pi_\gamma+R_\nu \pi_\nu+R_\gamma \pi_{\rm ex})=0~,\label{2}\\
&&\z^\p+\hub\z+\frac{1}{2k}\kappa\rho a^2 (R_\gamma\Delta_\gamma+R_\nu\Delta_\nu+R_b\Delta_b+R_c\Delta_c+R_\gamma\Delta_{\rm ex}\nonumber\\
&&+3(R_\gamma\delta P_\gamma+R_\nu\delta P_\nu+R_b\delta P_b+R_\gamma \delta P_{\rm ex}))=0~,\label{3}\\
&&\frac{2}{3}(\z-\s)+\frac{1}{k^2}\kappa\rho a^2(R_\gamma q_\gamma+R_\nu q_\nu+R_b q_b+R_\gamma q_{\rm ex})=0~,\label{4}\\
&&2\Phi-\frac{1}{k^2}\kappa\rho a^2(R_\gamma\Delta_\gamma+R_\nu\Delta_\nu+R_b\Delta_b+R_c\Delta_c+R_\gamma\Delta_{\rm ex}+R_\gamma\pi_\gamma+R_\nu\pi_\nu+R_\gamma\pi_{\rm ex})\nonumber\\
&&-\frac{3\hub}{k^3}\kappa\rho a^2(R_\gamma q_\gamma+R_\nu q_\nu+R_b q_b+R_\gamma q_{\rm ex})=0~.\label{5}
\end{eqnarray}
Here, the prime denotes a derivative with respect to 
conformal time $\tau$ and $\hub\equiv a^\p/a$.
We have defined an equation of state  parameter $\tilde{\gamma}$ as $p=(\tilde{\gamma}-1)\rho$, where
$p$ is the total pressure and $\rho$ is total energy density, and 
the gravitational constant is written $\kappa\equiv 8\pi G$.
The total energy density $\rho$ is
written $\rho=\rho_\gamma+\rho_\nu+\rho_b+\rho_c$, and 
 we have introduced the energy fractions  $R$, e.g.  $R_\gamma=\rho_\gamma/\rho$,
$R_\nu=\rho_\nu/\rho$, $R_b=\rho_b/\rho$ and $R_c=\rho_c/\rho$.  
The subscript denotes each component, i.e.  the photon ($\gamma$),
neutrino ($\nu$), baryon ($b$), CDM ($c$) and the extra source (${\rm ex}$) contributions. 
For normal fluid variables, the density fluctuations $\Delta$, the heat flux $q$,
 and the pressure fluctuations $\delta P$ are normalized using their background fluid energy density, e.g. 
$  \rho_\gamma \Delta_\gamma = \delta \rho_\gamma$, 
$ \rho_\nu \Delta_\nu = \delta \rho_\nu$, etc. 
For the fluctuations related to the extra source  $\Delta_{\rm ex}$, $q_{\rm ex}$ and $\pi_{\rm ex}$,   however, they enter the energy-momentum tensor directly without an associated background fluid energy density.  
Since we assume that they scale as $a^{-4}$, we can conveniently normalize  to the photon energy density.
Hence, in analogy with the other fluctuations,  in the Fourier space, we define $\Delta_{\rm ex}$, $q_{\rm ex}$, and $\pi_{\rm ex}$ by 
\begin{eqnarray}
&&\rho_\gamma\Delta_{\rm ex}= -T^{~~0}_{{\rm ex}~0}(k)~,\\
&&\rho_\gamma q_{\rm ex}=\hat{k}^iT^{~~0}_{{\rm ex}~i}(k)~,\\
&&\rho_\gamma\pi_{\rm ex}=-\frac{3}{2}(\hat{k}_i\hat{k}^j-\frac{1}{3}\delta_i^{~j})T^{~~i}_{{\rm ex}~j}(k)~,
\end{eqnarray}
where $\hat{k}_i$ is the wave vector with a unit length and $T^{~~\mu}_{{\rm ex}~\nu}$ is the energy-momentum tensor 
for the extra stress.
Note that, because we write the extra ansiotropic stress as $\rho_\gamma\pi_{\rm ex}$,  $\rho_\gamma$ carries the required  $a^{-4}$ scaling.  Therefore, $\pi_{\rm ex}$ becomes a fixed normalization constant. 
As noted above, the spectra of $\Delta_{\rm ex}$, $q_{\rm ex}$ and $\pi_{\rm ex}$ are not given by the Einstein equation and require a specific model of the extra dimensions.

Well into the radiation dominated epoch, the background energy densities for 
the baryons and CDM are negligible. 
Here, in this epoch, we take 
$\rho\simeq\rho_\gamma+\rho_\nu$,  $R_\gamma+R_\nu\simeq1$ and the $R_\gamma$ and $R_\nu$ are constant. 
Since we focus on the early universe, we need to study the
evolution of perturbations on super horizon scales, i.e. $k\tau\ll 1$.

The curvature perturbation $\eta$ is defined as
\begin{eqnarray}
\eta\equiv -(2\Phi+\s^\p/k)~.\label{6}
\end{eqnarray}
$\eta$ also satisfies the relation $\eta^\p=-\frac{k}{3}(\z-\s)$ \cite{2002PhRvD..65j4012L}.

Using the fact of $\kappa\rho a^2=3\hub^2$ and $\kappa(3\tilde{\gamma}-2)\rho a^2=-6\hub^\p$, we
can rewrite Eqs. (\ref{1})-(\ref{5}) as follows:
\begin{eqnarray}
&&\Phi^\p+\hub\Phi+\frac{1}{k}(\hub^2-\hub^\p)\sigma+\frac{3}{2k}\hub^2(R_\gamma q_\gamma+R_\nu q_\nu+ R_b q_b+R_\gamma q_{\rm ex})\nonumber\\
&&+\frac{3\hub}{2k^2}(\hub^2-2\hub^\p)(R_\gamma\pi_\gamma+R_\nu\pi_\nu+R_\gamma\pi_{\rm ex})-\frac{3\hub^2}{2k^2}(R_\gamma\pi_\gamma^\p+R_\nu\pi_\nu^\p+R_\gamma\pi_{\rm ex}^\p)\nonumber\\
&&=0~,\label{new1}\\
&&\sigma^\p+\hub\sigma+k\Phi+\frac{3\hub^2}{2k}(R_\gamma \pi_\gamma+R_\nu \pi_\nu+R_\gamma \pi_{\rm ex})=0~,\label{new2}\\
&&\z^\p+\hub\z+\frac{3\hub^2}{2k}(R_\gamma\Delta_\gamma+R_\nu\Delta_\nu+R_b\Delta_b+R_c\Delta_c+R_\gamma\Delta_{\rm ex}\nonumber\\
&&+3(R_\gamma\delta P_\gamma+R_\nu\delta P_\nu+R_b\delta P_b+R_\gamma \delta P_{\rm ex}))=0~,\label{new3}\\
&&\frac{2}{3}(\z-\s)+\frac{3\hub^2}{k^2}(R_\gamma q_\gamma+R_\nu q_\nu+R_b q_b+R_\gamma q_{\rm ex})=0~,\label{new4}\\
&&2\Phi-\frac{3\hub^2}{k^2}(R_\gamma\Delta_\gamma+R_\nu\Delta_\nu+R_b\Delta_b+R_c\Delta_c+R_\gamma\Delta_{\rm ex}+R_\gamma\pi_\gamma+R_\nu\pi_\nu+R_\gamma\pi_{\rm ex})\nonumber\\
&&-\frac{9\hub^3}{k^3}(R_\gamma q_\gamma+R_\nu q_\nu+R_b q_b+R_\gamma q_{\rm ex})=0~.\label{new5}
\end{eqnarray}
Hereafter, we neglect the matter contribution because we are in deep radiation era.
We can also neglect photon anisotropic stress since we focus on the epoch
before recommbination.

Eliminating $q$'s from Eqs. (\ref{new4}) and (\ref{new5}), we obtain following equation:
\begin{eqnarray}
&&2\Phi-\frac{3\hub^2}{k^2}(R_\gamma\Delta_\gamma+R_\nu\Delta_\nu+R_\gamma\Delta_{\rm ex}+R_\nu\pi_\nu+R_\gamma\pi_{\rm ex})\nonumber\\
&&+\frac{2\hub}{k}(\z-\s)=0\label{nn5}~.
\end{eqnarray}
In this paper, we assume an adiabatic condition.
In this case, $\delta P$ is expressed as $\delta P = c_s^2\Delta$, where $c_s^2$ is the adiabatic sound speed. 
Using the relations  $\Delta_\gamma=3\delta P_\gamma$, $\Delta_\nu=3\delta P_\nu$ and $\Delta_{\rm ex}=3\delta P_{\rm ex}$, Eq. (\ref{new3}) becomes:
\begin{eqnarray}
\z^\p+\hub \z+\frac{3\hub^2}{k}(R_\gamma\Delta_\gamma+R_\nu\Delta_\nu+R_\gamma\Delta_{\rm ex})=0\label{nn3}
\end{eqnarray}
From Eqs.(\ref{nn5}) and (\ref{nn3}), we obtain
\begin{eqnarray}
\z^\p +3\hub\z+2k\Phi-2\hub\s=\frac{3\hub^2}{k}(R_\nu\pi_\nu+R_\gamma\pi_{\rm ex})\label{nnn3}
\end{eqnarray}
Next we eliminate $\Phi$ from Eqs. (\ref{new2}) and (\ref{nnn3}) using Eq.(\ref{6}):
\begin{eqnarray}
&&\s^\p+2\hub\s-k\eta=-\frac{3\hub^2}{k}(R_\nu\pi_\nu+R_\gamma\pi_{\rm ex})\label{nn2}\\
&&\z^\p +3\hub\z-k\eta-(\s^\p+2\hub\s)=\frac{3\hub^2}{k}(R_\nu\pi_\nu+R_\gamma\pi_{\rm ex})\label{nnnn3}
\end{eqnarray}
From Eqs.(\ref{nn2}) and (\ref{nnnn3}), we obtain
\begin{eqnarray}
\z^\p+3\hub\z-2k\eta=0\label{nnn2}
\end{eqnarray}
Eqs. (\ref{nn2}) and (\ref{nnn2}) together imply
\begin{eqnarray}
3\eta^\pp+9\hub\eta^\p-k\hub\s+k^2\eta=-3\hub^2(R_\nu\pi_\nu+R_\gamma\pi_{\rm ex})~~,\label{nnnn2}
\end{eqnarray}
where we have used the relation $\eta^\p=-\frac{k}{3}(\z-\s)$.
After dividing Eq.(\ref{nnnn2}) by $\hub$, we take the derivative of Eq.(\ref{nnnn2}) with $\tau$:
\begin{eqnarray}
&&3\hub^{-1}\eta^\ppp-3\hub^{-2}\hub^\p\eta^\pp+9\eta^\pp+k^2\hub^{-1}\eta^\p-k^2\hub^{-2}\hub^\p\eta-k\s^\p\nonumber\\
&&=-3\hub^\p(R_\nu\pi_\nu+R_\gamma\pi_{\rm ex})-3\hub(R_\nu\pi_\nu^\p+R_\gamma\pi_{\rm ex}^\p) ~~.
\label{etaeq}
\end{eqnarray}
Finally we obtain a third order equation for $\eta$ from Eqs.(\ref{nn2}), (\ref{nnnn2}) and (\ref{etaeq}):
\begin{eqnarray}
&&3\hub^{-1}\eta^\ppp+(15-3\hub^{-2}\hub^\p)\eta^\pp+(18\hub+k^2\hub^{-1})\eta^\p+k^2(1-\hub^{-2}\hub^\p)\eta\nonumber\\
&&=(-9\hub^2-3\hub^\p)(R_\nu\pi_\nu+R_\gamma\pi_{\rm ex})-3\hub (R_\nu\pi_\nu^\p+R_\gamma\pi_{\rm ex}^\p)~~.
\label{eta1}
\end{eqnarray}
In the radiation dominated epoch, $\hub$ is approximately $\tau^{-1}$.
Thus, we obtain the final result:
\begin{eqnarray}
&&\tau^3\eta^\ppp+6\tau^2\eta^\pp+(6+\frac{k^2\tau^2}{3})\tau\eta^\p+\frac{2}{3}k^2\tau^2\eta\nonumber\\
&&=-2(R_\nu\pi_\nu+R_\gamma\pi_{\rm ex})-\tau(R_\nu\pi_\nu^\p+R_\gamma\pi_{\rm ex}^\p)~.\label{eta2}
\end{eqnarray}
This is the equation for $\eta$ sourced by the anisotropic stress.

In order to solve the growth of the neutrino anisotropic stress, we need the relevant
Boltzmann equation \cite{1995ApJ...455....7M,2002PhRvD..65j4012L,1992ApJ...395...54D,1999ApJ...513....1C}:
\begin{eqnarray}
\pi_\nu^\p=k(\frac{2}{5}q_\nu-\frac{3}{5}G^{(3)}_\nu)+\frac{8}{15}k\sigma~,\label{bol2}
\end{eqnarray}
where $G_\nu^{(3)}$ is the octupole moment.
Eq. (\ref{bol2}) is valid after the neutrino decoupling.
Eliminating $\sigma$ from Eq.(\ref{nn2}) with Eq.(\ref{bol2}),
we obtain the following equation:
\begin{eqnarray}
\frac{d^2{\pi}_{\nu}}{d (\ln \tau)^2}+\frac{d{\pi}_{\nu}}{d\ln \tau}+\frac{8}{5}R_{\nu}\pi_{\nu}+{\cal O}(k\tau)=-\frac{8}{5}R_{\gamma}\pi_{\rm ex}~~,
\label{eq12}
\end{eqnarray}
where the term of ${\cal O}(k\tau)$ is negligible on super horizon scales.

\section{Analytical Solutions}

\begin{figure}
\includegraphics[angle=270,scale=0.3]{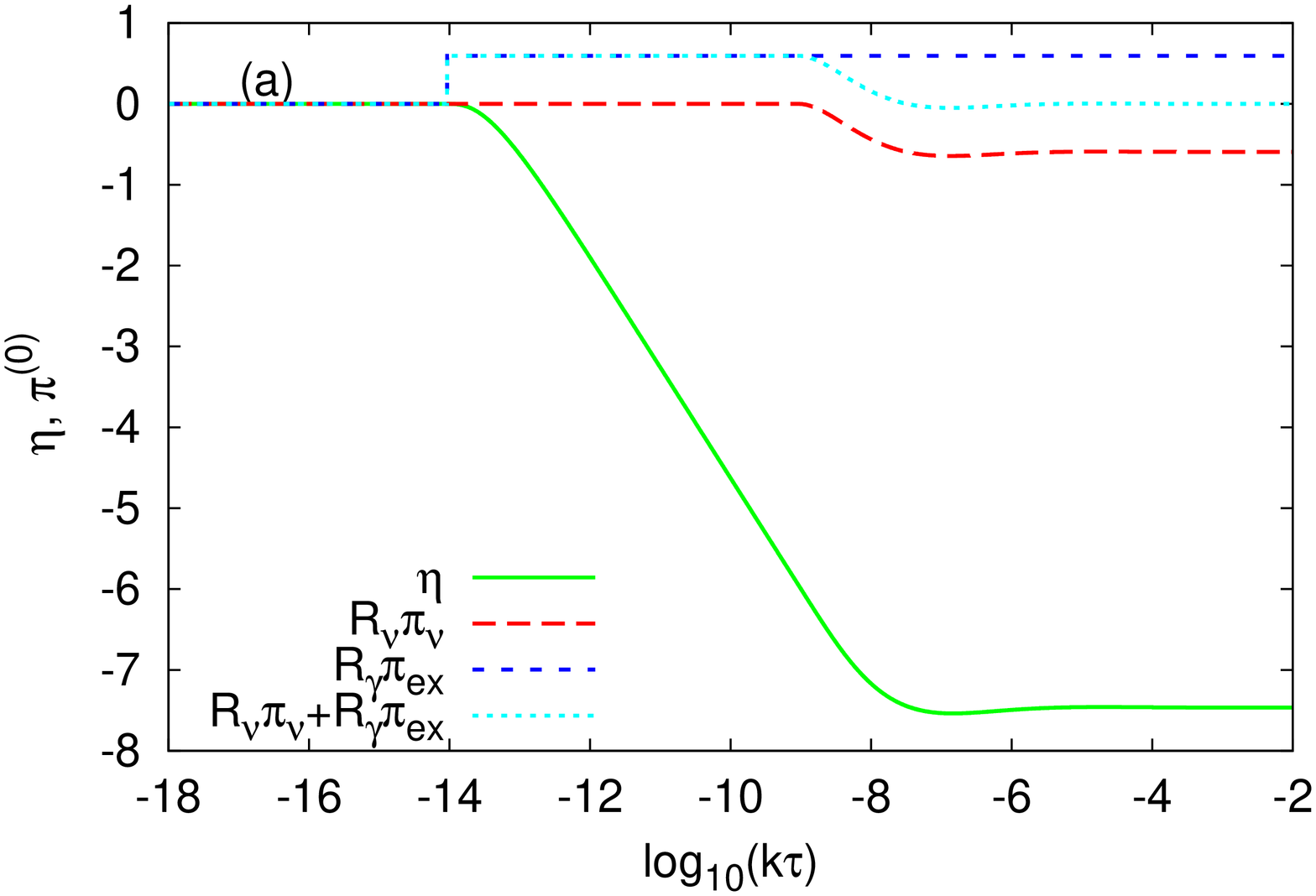}
\includegraphics[angle=270,scale=0.3]{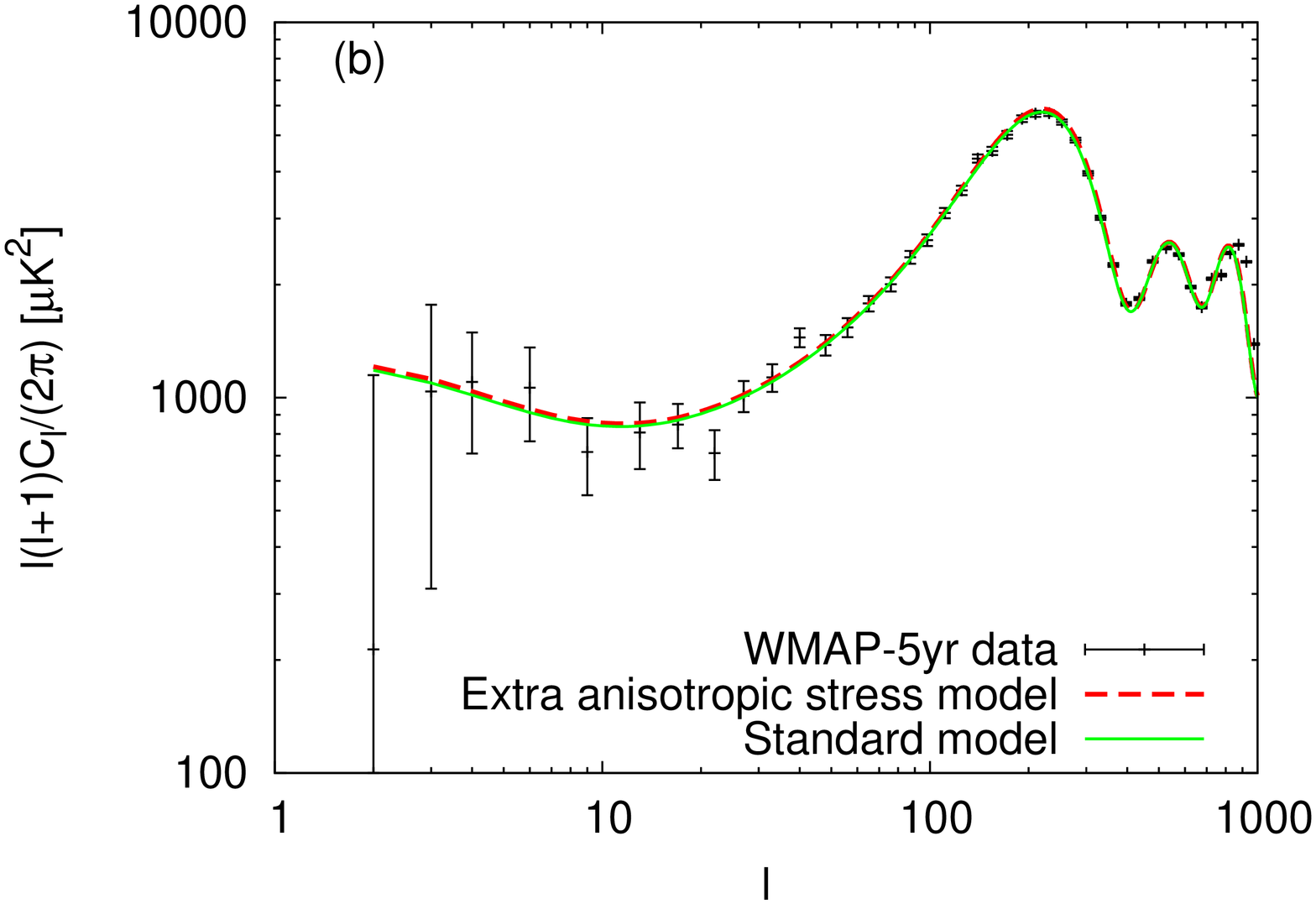}
\caption{(a) Numerical evolution of $\eta$, $R_{\nu}\pi_\nu$, $R_\gamma\pi_{\rm ex}$, and $R_\nu\pi_\nu+R_\gamma\pi_{\rm ex}$.
 Neutrino decoupling occurs at $\tau_\nu\sim9.3\times10^{-5}{\rm Mpc}$. We set $k=10^{-5}{\rm Mpc^{-1}}$, $r_\tau=10^5$
and $\pi_{\rm ex}=1~(\tau_{\rm in}\leq\tau)$
and assume for convenience that the extra anisotropic stress was  generated as a step function at $\tau_{\rm in}=\tau_\nu/r_\tau$.   
The neutrino anisotropic stress (red curve) cancels the extra anisotropic stress (violet curve). 
The curvature perturbation $\eta$ (green curve) stops growing after this cancellation.
Panel (b) shows fit to the WMAP-5yr data of the primary scalar-mode and passive scalar mode power spectra induced by the extra anisotropic stress in the TT mode. This fit is compared with the standard $\Lambda$CDM model. 
The two power spectra are nearly indistinguishable. Note however that we adopted  the spectral index which best fits the WMAP-5yr data \cite{2008arXiv0803.0547K} for  the extra anisotropic stress. \label{fig1}}
\end{figure}
To clarify the numerical evolution of the anisotropic stress we here describe an analytic model.
We can set $\pi_\nu\sim{\pi}_\nu^{\prime}\sim 0$ during the interval between the generation epoch of the extra anisotropic stress, $\tau_{\rm in}$, and the epoch of neutrino decoupling, $\tau_{\nu}$, i.e. $\tau_{\rm in}\leq\tau\leq\tau_\nu$.

First, we consider the case of $\tau_{\rm in}<\tau<\tau_\nu$.
Assuming $\pi_{\rm ex}\propto {\rm const.}$, 
we find the following physical homogeneous and inhomogeneous 
solutions for Eq. (\ref{eta2}),
\begin{eqnarray}
&&\eta\propto \tau^{-1},~const.~{\rm (homogeneous~solutions)},\nonumber\\
&&\eta= -R_\gamma\pi_{\rm ex}\ln \tau~{\rm (inhomogeneous~solution)}.
\end{eqnarray}
 Although the homogeneous solution $\eta\propto \tau^{-1}$ is a decaying mode,
we need this mode to calculate the growth of the curvature perturbations.
Finally we obtain
\begin{eqnarray}
\eta(\tau)-\eta(\tau_{\rm in})= -R_{\gamma}\pi_{\rm ex}\left(\ln (\tau/\tau_{\rm in})+\tau_{\rm in}/\tau-1\right)~,
\end{eqnarray}
where we have used the initial condition $\eta=\eta(\tau_{\rm in})$, $\eta^\p=0$ at $\tau=\tau_{\rm in}$.
In this epoch, the curvature perturbation grows
logarithmically.

After neutrino decoupling, $\tau\geq\tau_\nu$, the
neutrino anisotropic stress must be evolved with the use of the Boltzmann equation.
We can solve Eq. (\ref{eq12}) analytically to obtain 
\begin{eqnarray}
\pi_\nu=c_1e^{(-1/2+i\sqrt[]{\mathstrut 32R_\nu/5-1})\ln\tau}+c_2e^{(-1/2-i\sqrt[]{\mathstrut 32R_\nu/5-1})\ln\tau}-\frac{R_\gamma}{R_\nu}\pi_{\rm ex}~,\label{sol}
\end{eqnarray}
where $c_1$ and $c_2$ are constants of integration.
Since the first two terms on the r.h.s. of Eq. (\ref{sol}) damp to  zero, 
 the neutrino anisotropic stress eventually cancels the extra anisotropic stress asymptotically,
\begin{eqnarray}
R_\nu\pi_{\nu} \rightarrow -R_{\gamma} \pi_{\rm ex}~.
\label{eq14}
\end{eqnarray}
Since $\eta$ satisfies $\eta^\p=-\frac{k}{3}(\z-\s)$, we can rewrite Eq.(\ref{bol2}) as follows:
\begin{eqnarray}
\pi_\nu^\p=\frac{8}{5}\eta^\p+\frac{8}{15}k\z+\frac{2}{5}kq_\nu-\frac{3}{5}kG_\nu^{(3)}~
\end{eqnarray}
Assuming that $k\z$, $kq_\nu$ and $kG_\nu^{(3)}$ are negligible on super horizon scales,
we have
\begin{eqnarray}
\eta(\tau)-\eta(\tau_\nu)\simeq\frac{5}{8}\int^\tau_{\tau_\nu}\pi_\nu^\p(\tilde{\tau}) d\tilde{\tau}~.
\label{eq14-2}
\end{eqnarray}
Thus, we obtain an asymptotic solution for  $\eta$ from Eqs. (\ref{eq14}) and (\ref{eq14-2}):
\begin{eqnarray}
\eta(\tau)-\eta(\tau_\nu) \rightarrow -\frac{5}{8}\frac{R_{\gamma}}{R_\nu} \pi_{\rm ex}~.
\end{eqnarray}
After the complete cancellation of the anisotropic stress, therefore, the curvature perturbation 
stays constant.

Finally, we obtain the result:
\begin{eqnarray}
&&\eta \rightarrow \eta_{pr}+\eta_{ps1}+\eta_{ps2}~,
\nonumber\\
&&\eta_{ps1}\equiv -R_{\gamma}\pi_{\rm ex}\left(\ln r_{\tau}+r_{\tau}^{-1}-1\right)~,
\nonumber\\
&&\eta_{ps2}\equiv -5R_{\gamma} \pi_{\rm ex}/(8R_{\nu})~~,
\label{eq15}
\end{eqnarray}
where $r_\tau\equiv \tau_\nu/\tau_{\rm in}$, 
$\eta_{pr}$ is the primary perturbation, and $\eta_{ps1}$ represents the perturbations which are generated by $\pi_{\rm ex}$ during the epoch of $\tau_{\rm in}\leq\tau\leq\tau_{\nu}$.  The quantity
$\eta_{ps2}$ is the  perturbation induced by the growth of 
the anisotropic stress after neutrino decoupling. 
This provides insight into the numerical solutions described in the next section.

\section{Numerical Solutions}
We solved Eqs. (\ref{eta2}) and (\ref{eq12}) numerically. 
Fig.~\ref{fig1}(a) shows an example
of  the numerical evolution of $\eta$ and the $\pi$'s. 
For purposes of illustration we in this figure have set $k=10^{-5}~{\rm Mpc^{-1}}$, $r_\tau=10^5$
and $\pi_{\rm ex}=1~(\tau_{\rm in}\leq\tau)$.  (Note that $r_\tau=10^5$ is not the same as the  value adopted for figure \ref{fig1}(b), but is chosen here to make the graph easier to read).  We have also
 assumed that the extra anisotropic stress was generated at $\tau_{\rm in}=\tau_\nu/r_\tau$  as a step function.  This choice of a step function is a matter of convenience.  
In a realistic model, the appearance of this term would be a result of dynamics in the bulk dimensions.  
Nevertheless, as should be apparent from this plot, 
it does not particularly matter how the anisotropic stress grows in as long as $\pi_{\rm ex}$ obtains its constant value sometime before neutrino decoupling.
The numerical calculation agrees well with our analytical
solutions [Eqs. (\ref{eq14}) and (\ref{eq15})].

This asymptotic behavior to obtain constant $\eta$ is very similar to the standard adiabatic mode 
whose $\eta$ also remains constant on super horizon scales. 
These primary-like modes which are induced by the anisotropic stress,
are called "passive modes", while the cancellation of the anisotropic stress generates
 "compensated modes" \cite{2004PhRvD..70d3011L}.

From the  above discussions, we can conclude that if $r_{\tau}$ is large enough, the passive scalar mode
 can mimic the primary mode. 
If there is a primordial stress source, $T^{\mu}_{~\nu}$, which does not contribute to the background and satisfies
 $T^{\mu}_{~\mu}=T^{\mu}_{~\nu;\mu}=0$, just like the dark radiation in the low-energy limit, the energy density and pressure grow as $\propto a^{-4}$ on superhorizon scales. Assuming that the anisotropic stress also scales as $\propto a^{-4}$, we 
can calculate the CMB spectrum induced by an extra anisotropic stress as shown in Fig. \ref{fig1}(b). 
Here we set $r_\tau=10^{18}$ corresponding to  the epoch at the end of inflation, 
and $\sqrt[]{\mathstrut \mid\pi_{\rm ex}\mid^2}\simeq 8.6\times 10^{-6}$ from 
a fit to  the observed amplitude of initial curvature perturbation $\Delta_{\cal R}^2=2.41\times 10^{-9}$ \cite{2008arXiv0803.0547K}.
In addition, we adopt $\Delta_{\rm ex}\simeq -\pi_{\rm ex}$ as in
the case of a primordial magnetic field \cite{2007PhRvD..75b3002K}
 and we have used the best fit parameters from the WMAP-5yr analysis adopting  the best fit spectral index for the extra anisotropic stress spectrum.  

As can be seen in Fig. \ref{fig1}(b), the CMB anisotropies induced by the extra anisotropic stress is very similar to
the primary spectrum induced by inflation. 
Note, however, that although we have generated a spectrum of the extra anisotropic stress to be consistent with the observation,
this requires that the magnitude and spectrum be of the correct form to mimic the standard inflation-generated fluctuations.
In general, the observed CMB spectrum could be a superposition of spectra generated by inflation (or some other mechanisms) and the anisotropic 
stress, as shown in Eq. (\ref{eq15}). 
Therefore, further study of the origin of the anisotropic stress and whether it can have the magnitude and spectrum consistent with the observed spectrum is
highly desirable.

The Gaussianity of the fluctuations in our model depends upon the source of the extra anisotropic stress. 
For example, a PMF spectrum is non-Gaussian.
Since a normalized anisotropic stress $\sqrt[]{\mathstrut \mid\pi_{\rm ex}\mid^2}\simeq 8.6\times 10^{-6}$ 
corresponds to about $9.3{\rm nG}$,
which is the same order as the PMF estimated in Ref.~\cite{2001PhR...348..163G},  
one should take account of the passive components in scalar and tensor modes of the CMB.  It might be  expected, therefore, that the whole observed spectrum of the present CMB  be explained by the passive scalar mode of the PMF.  However, that is not the case 
 because the tensor-to-scalar ratio of the passive PMF mode turns out to be greater than unity from previous work \cite{2008arXiv0811.0230P}.  Therefore a PMF  is excluded by the WMAP-5 data.
There is also another difficulty in that the spectrum of the PMF is not well known.

We can understand the growth of curvature perturbations from the Einstein equation.  The traceless part of the Einstein tensor is equal to
the anisotropic stress in the universe, cf. Eq. (\ref{2}).
Since the traceless part of the Einstein tensor is written in terms of 
the anisotropic expansion rate of the universe, $\s$, 
the anisotropic stress causes the universe to expand anisotropically. 
As the anisotropy of the universe increases, the curvature of the universe also
increases.
After neutrino decoupling, the neutrino anisotropic stress grows.  This tends  to decrease the
anisotropic expansion rate, Eq. (\ref{bol2}), i.e.  neutrinos cancel the extra anisotropic stress.
After this compensation, the anisotropic expansion rate vanishes and the universe
expands isotropically.
In this phase, the curvature perturbation, $\eta$, is conserved.

\section{Conclusion}

We have found a simple solution for Eq. (\ref{eta2}) and have
shown that if there exists an extra anisotropic stress which scales as $\rho_\gamma\pi_{\rm ex}\propto a^{-4}$, the anisotropic stress from neutrinos $\pi_\nu$ exactly cancels $\pi_{\rm ex}$. 
Before neutrino decoupling, the curvature perturbations grow logarithmically.
After neutrino decoupling, they become constant on superhorizon scales 
just like the standard adiabatic mode of inflation.  
This is  because the 
total anisotropic stress vanishes via a  cancellation.
Thus, the resultant CMB spectrum is a superposition of the primary  and passive modes.

As an illustration we have considered the possibility that the passive scalar mode has a scale invariant spectrum.  In this case  the extra anisotropic stress might even produce a power spectrum similar to the observed CMB. 
This suggests a possible additional consideration in the determination of cosmological parameters from the CMB.  
Also we note that the  Gaussianity of the CMB fluctuations depends upon the  source for the extra anisotropic stress.  
Hence, should this extra anisotropic stress be present in the observed power spectrum,
 it might be detectable by non-Gaussianity in the fluctuations.  
Future CMB observations of  non-Gaussianity may, therefore,  help to constrain this possibility.

To summarize, our purpose here has been to suggest that such a contribution to the observed spectrum may exist. 
Thus, further studies on the amplitude and spectrum of the extra anisotropic stress in brane-world cosmology are warranted.

\acknowledgments
In this work, we have used the modified CAMB code of Ref. \cite{Lewis:1999bs}.
This work was supported in part by Grants-in-Aid for Scientific
Research (20244035, 20105004) of the Ministry of Education, Culture, Sports,
Science and Technology of Japan. Work at UND was supported in part by the US Department of Energy under research grant DE-FG02-95-ER40934.
\bibliography{astress}

\clearpage

\end{document}